# Rencontre improbable entre von Foerster et Snowden
## L'éclairage de la seconde cybernétique sur la révolution du Big Data

### 1. Introduction

En 1976 à Cuernavaca, Heinz von Foerster, fondateur de la seconde cybernétique et précurseur dans le domaine des systèmes complexes, intervient dans un séminaire d'Ivan Illich, penseur de l'écologie politique. Suite à l'analyse par ce dernier de sa notion de *contre-productivité* – auto-dérégulation et auto-désorganisation d'un système qui devient étranger aux éléments qui le constituent[i] – von Foerster émet une conjecture au caractère visionnaire :

> « Ce que vous essayez de décrire, c'est le rapport de causalité circulaire entre une totalité (par exemple, une collectivité humaine) et ses éléments (les individus qui la composent). Les individus sont liés les uns aux autres, d'une part, ils sont liés à la totalité, d'autre part. Les liens entre individus peuvent être plus ou moins « rigides » - le terme technique que j'emploie est « triviaux ». Plus ils sont triviaux, moins, par définition, la connaissance du comportement de l'un d'eux apporte d'information à l'observateur qui connaît déjà le comportement des autres. Je conjecture la relation suivante : plus les relations inter-individuelles sont triviales, plus le comportement de la totalité apparaîtra aux éléments individuels qui la composent comme dotée d'une dynamique propre qui échappe à leur maîtrise.
>
> Je conçois que cette conjecture présente un caractère paradoxal, mais il faut bien comprendre qu'elle n'a de sens que parce que l'on prend ici le point de vue, intérieur au système, des éléments sur la totalité. Pour un observateur extérieur au système, il va de soi que la trivialité des relations entre éléments est au contraire propice à une maîtrise conceptuelle, sous forme de modélisation. Lorsque les individus sont trivialement couplés (du fait de comportement mimétique, par exemple) la dynamique du système est prévisible, mais les individus se sentent impuissants à en orienter ou réorienter la course, alors même que le comportement d'ensemble continue de n'être que la composition des réactions individuelles à la prévision de ce même comportement. Le tout semble s'autonomiser par rapport à ses conditions d'émergence et son évolution se figer en destin. ».

Cette proposition fut rapportée par Jean-Pierre Dupuy sous l'appellation de « conjecture de von Foerster » (Dupuy 2006). Il lui conféra en 1987 le statut de théorème dans le cadre de la théorie de l'information lors d'une collaboration avec Moshe Koppel et Henri Atlan (Koppel et al. 1987).

Trois ans après 1984, qui n'avait pas délivré ses promesses littéraires, peu de gens étaient enclins à accepter l'idée qu'un théorème mathématique puisse rendre compte un tant soit peu des phénomènes sociaux. Aujourd'hui encore, certains affirmeront que de telles généralités sur le social ne peuvent être fondées, ne serait-ce que parce que la notion d'expérimentation à l'échelle d'une société reste en soi problématique. Nous allons pourtant montrer que les technologies de l'information et de la communication (TIC), devenues ubiquitaires dans nos sociétés, nous donnent à la fois un exemple de ce que von Foerster appelait « relations rigides », un terrain d'expérimentation et une validation empirique de l'intuition de von Foerster. Les implications de cette conjecture sont nombreuses, et permettent notamment d'apporter un éclairage original sur les récentes révélations d'Edward Snowden, qui constituent une démonstration incontestable du fait que l'accès à un point de vue « extérieur » à nos sociétés numériques est devenu un enjeu stratégique pour de nombreux acteurs.

### 2. Les urnes de Polya et le destin illusoire

La conjecture de von Foerster mériterait sans doute d'être contextualisée par un exposé détaillé des concepts fondateurs de la seconde cybernétique et des approches « systèmes complexes » : auto-organisation, auto-référence, rétroactions positives/négatives, etc. Par souci de concision, nous nous limiterons cependant aux exemples minimaux requis pour sa compréhension, renvoyant le lecteur souhaitant acquérir une profondeur de champ à l'ouvrage fondateur (Dumouchel 1983), ainsi qu'à certaines publications récentes (Bourgine et al. 2008, Clarke and Hansen 2009) .

Commençons par une expérience de pensée volontairement schématique. Imaginons que deux réseaux sociaux *A* et *B* – tels que Facebook et Google+ – fassent leur entrée sur le marché le même jour avec des fonctionnalités identiques et un seul utilisateur pour chacune. Imaginons de plus que chaque plate-forme acquiert de nouveaux utilisateurs de la manière suivante : régulièrement un utilisateur de l'une de ces deux plates-formes décide d'inviter une personne qui ne connaît pas encore les réseaux sociaux à venir s'y inscrire pour y devenir son « ami », celle-ci acceptant toujours l'invitation. Si à tout moment, tout utilisateur, quelque soit la plate-forme utilisée, a les mêmes chances de lancer une invitation que se passera-t-il ? A la première invitation, issue par exemple de *A*, il y aura deux utilisateurs pour *A* et un seul pour *B*. Par la suite, il y aura donc deux fois plus de chances pour qu'une invitation émane d'un utilisateur de *A* plutôt que de *B*. Et après ?

Comme le montre la figure 1, la simulation informatique de ce processus montre que la part de marché de la plate-forme *A* évolue dans un premier temps de manière assez imprévisible pour venir se stabiliser autour de 80 %. Cette convergence vers une répartition stable des parts de marché semble inéluctable (il y a de moins en moins de fluctuations), malgré le caractère aléatoire des initiatives des utilisateurs, qui peuvent appartenir à l'une ou l'autre des plates-formes. Un analyste serait alors tenté d'interpréter *a posteriori* l'écart de popularité entre les deux réseaux sociaux comme reflétant une valeur intrinsèque des deux plates-formes, puisque après un certain temps, rien ne semble pouvoir changer l'équilibre des forces malgré un flot constant de nouveaux utilisateurs. Il serait pourtant dans l'erreur, les plates-formes étant strictement équivalentes.

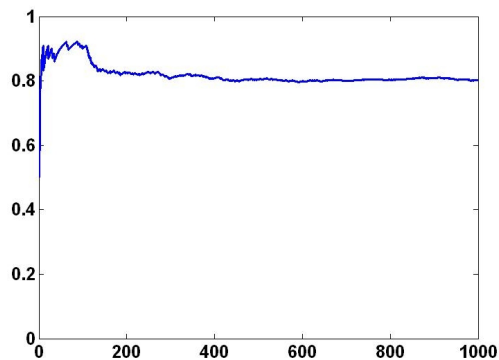

*Figure 1. Simulation de l'évolution des parts de marché de la plate-forme A après 1000 inscriptions.*

Le processus que nous venons de décrire est connu en mathématiques sous le nom d'Urnes de Polya (Johnson et al. 1977). Sa dynamique peut être décrite très précisément. En reprenant la sémantique utilisée ci-dessus, il est démontré que la répartition finale des parts de marché des deux plates-formes est tout à fait contingente : une autre expérience convergerait vers une valeur différente. Mieux, toutes les valeurs entre 0 % et 100 % de parts de marché pour la plate-forme *A* ont les mêmes chances d'être observées.

Ce phénomène est une notion clé de l'étude des systèmes dynamiques appelée *dépendance au chemin (path dependancy)* : bien que la règle d'évolution du système ne change pas au cours du temps, l'ensemble des possibles pour les états du système – la distribution des parts de marché dans notre cas – se rétrécit avec le temps. Au cours d'une expérience, du point de vue des utilisateurs (intérieur au système), il est ainsi possible d'avoir l'illusion qu'une raison objective guide son évolution vers une répartition déterminée des parts de marché. Cependant, un observateur extérieur, connaissant le processus d'inscription, n'y verrait que de la contingence stabilisée par des rétroactions positives[1].

En termes plus techniques, la proportion asymptotique des parts de marché est appelée *attracteur* et un système modélisable par des Urnes de Pólya a la particularité d'avoir une infinité d'attracteurs équiprobables.

Cette expérience abstraite, aussi simple soit-elle, illustre quelques concepts importants de la seconde

---

[1] On dit qu'il y a rétroactions positives lorsqu'on se trouve en présence d'effets qui renforcent leurs causes.

cybernétique :

- Les systèmes dynamiques peuvent avoir un nombre arbitrairement grand d'attracteurs stables, en particulier en présence de rétroactions positives induisant des dépendances au chemin,
- La sélection de l'un de ces attracteurs au cours du temps peut être totalement contingente, initiée par des fluctuations aléatoires,
- L'interprétation du processus de sélection de l'attracteur dépend de l'observateur. En particulier, vu de l'intérieur, par exemple par un utilisateur de l'une des plateformes, ce processus pourra être interprété comme l'émergence d'un ordre auto-organisé (distribution stable des parts de marché). En revanche, vue de l'extérieur, *i.e.* par un observateur n'interférant par avec la dynamique du système, il pourra être interprétée comme la sélection contingente de l'un de ses attracteurs.

un principe présent dans un grand nombre situations réelles : une *dépendance au chemin* engendrée par des *rétroactions positives*.

**3. L'ère de la recommandation**

L'expérience de pensée précédente permet d'appréhender une notion importante de la conjecture de von Foerster, celle de relations « rigides » ou « triviales » : l'action de chaque agent (s'inscrire comme nouvel utilisateur) est déterminée de manière univoque par le choix de l'une de ses connaissances. Nous pourrions dire que c'est le système de recommandation le plus trivial. Or, il existe un phénomène sociétal extrêmement récent, inimaginable à l'époque de la formulation de la conjecture de von Foerster, et puissant facteur de « trivialité » dans les relations humaines : la surcouche de systèmes de recommandation et de classement (*ranking*) qui recouvre progressivement le Web. Leur finalité est précisément de suggérer des actions ou des idées à un internaute en fonction de l'historique de ses actions (duquel sont inférées ses préférences), de ceux des autres utilisateurs et, plus récemment, de l'historique des actions des membres de son réseau social.

Rappelons que Google, qui capte plus de 65 % des recherches d'information dans le monde et est devenue en 15 ans l'une des premières entreprises mondiale, a fait sa notoriété sur l'algorithme pageRank, dont l'hypothèse est que la meilleure manière de trouver une information sur le web est d'utiliser les liens laissés par les internautes pour recommander des sites. Mais les systèmes de recommandation ont fait le succès d'autres géants de l'Internet comme Amazon ou Netflix, qui proposent à leurs clients des produits qu'ils vont probablement aimer en fonction de leur profil d'achat, et plus récemment des profils d'achat de leurs amis. La performance de ces systèmes de recommandation et leur capacité prédictive sur le comportement des internautes n'est plus à prouver et l'utilisation d'un réseau social en ligne est devenue la panacée de ces pratiques. Aucun site d'e-commerce, aucune plate-forme de réseau social ne s'aventurerait aujourd'hui sans un système de recommandation de nouveaux amis, de nouveaux produits ou un système de filtrage collaboratif. C'est une composante si stratégique pour ces acteurs du Web, qu'une entreprise comme Netflix, qui propose des films en téléchargement à quelques 25 millions d'abonnés, n'a pas hésité à proposer 1.000.000$ à toute personne qui réussirait à améliorer de 10% les performances prédictives de son système de recommandation. Pour donner une idée de la « rigidité » introduite par de tels systèmes, il faut savoir que 70% des internautes font confiance aux recommandations en ligne émanant d'inconnus, ce taux de confiance grimpant à 90 % lorsqu'ils connaissent la personne[ii]. Pour un site tel que Netflix, 60 % des achats sont consécutifs à une recommandation.

Il est important de remarquer que dans le cas de Netflix, il ne s'agit pas d'une recommandation sur des produits de consommation classiques, pour lesquels nous pourrions imaginer qu'elle ne fait que faciliter la mise à jour d'une information objective, mais de produits culturels, dont l'appréciation est censée être relativement subjective. On peut alors se poser la question de l'impact des systèmes de recommandation ou de classement sur les dynamiques sociales dans des domaines aussi divers que le choix de biens culturels (littérature, musique, cinéma), l'évolution des sciences (via la recommandation de travaux antérieurs), le choix de destinations touristiques, les réseaux d'amitié et la mixité sociale via les réseaux sociaux et les sites de rencontres (voire plus si affinités) ou les dynamiques d'opinion (notamment en politique où les effets de la manipulation des moteurs de recherche a été documentée (Epstein and Robertson 2014) et où on relèvera l'apparition de plate-formes utilisant le principe des réseaux sociaux en ligne pour organiser le débat citoyen). Dans un monde qui croule sous les masses de données, les systèmes de recommandation et de classement sont une béquille indispensable dont l'impact culturel sera peut-être un jour plus important que celui des initiatives étatiques dans le domaine de l'éducation.

Lors d'une récente conférence TED [SLA 11], Kevin Slavin défendait l'idée de l'apparition d'un nouvel acteur dans la co-évolution entre l'Humanité et la Nature : les algorithmes. Il y démontre de manière convaincante que le choix des algorithmes dans les différentes sphères des activités humaines ont des effet sociétaux et environnementaux mesurables et significatifs. Avec la conjecture de von Foerster en arrière-plan, nous pouvons effectivement poser la question de l'impact sur les sociétés humaines et leur gouvernance, des algorithmes de recommandation et de classement, et plus généralement des TIC, dans la mesure où celles-ci changent massivement la manière dont les gens s'influencent mutuellement dans leurs goûts et leurs actions.

Dans la sphère académique, les sciences de la gestion et du marketing ont été les premières, comme on peut le comprendre, à investir le domaine de la recommandation. Une des questions que se sont initialement posées ces communautés est de savoir si les systèmes de recommandation que l'on trouve sur le Web favorisent la diversification de la consommation ou au contraire l'émergence de produits stars et de blockbusters. Les études empiriques ont apporté différentes réponses à cette question (Felder and Hosanagar 2009 ; Zhou et al. 2011 ; Brynjolfsson et al. 2007) et on ne s'étonnera pas d'apprendre que des approches formelles, s'inspirant justement des urnes de Polya, ont pu montrer que les deux cas de figures sont possibles en fonction du paramétrage des algorithmes utilisés (Felder and Hosanagar 2009). Le choix du type d'algorithme, *in fine* de la diversification des choix des utilisateurs et indirectement de leurs goûts (y compris en ce qui concerne le type de biens dits « culturels ») est alors assujetti à la stratégie commerciale de l'entreprise. En témoigne cette préconisation issue d'un article de gestion, après une analyse de différents retours d'expérience : « si les gens achètent vos produits à l'unité, choisissez un système de recommandation conservateur [qui tend à réduire la diversité] ; s'ils vous aiment assez pour payer un abonnement mensuel, ils seront probablement ouverts à des systèmes de recommandation qui leur réserveront de plaisantes surprises » (Davenport et Harris 2009). L'étendue de la culture générale de la population d'un pays dépendra-t-elle un jour autant de la stratégie commerciale des grands distributeurs que des choix des ses institutions en matière d'éducation ?

**4. Validation empirique de la conjecture de von Foerster**

Quelques articles à visées plus larges sortent du champ des sciences de gestion et apportent une perspective tout à fait intéressante sur les effets des couplages inter-individuel sur les dynamiques sociales. Un article paru dans *Science (Salganick et al. 2006)* conçoit ainsi l'une des premières expériences à grande échelle étudiant l'effet de l'influence sociale sur les dynamiques culturelles. Leur point de départ est le constat que les blockbusters dans le domaine du cinéma, de la littérature ou de la musique inspirent souvent aux analystes l'idée qu'ils sont qualitativement différents du reste de la production, malgré le fait que les experts échouent régulièrement à en prédire le succès.

Pour mieux comprendre ce paradoxe, les auteurs ont créé un marché musical artificiel auquel ont pris part plus de 14.000 participants. Ceux-ci ont pu télécharger sur une plate-forme mise en place pour l'expérience des titres musicaux de groupes jusque là inconnus. Chaque participant avait la possibilité de choisir un titre musical, de l'écouter, de le noter sur une échelle de un à cinq, puis de le télécharger. Les participants ont été répartis suivants des conditions expérimentales différentes caractérisées par la présence ou l'absence d'information sur le nombre de téléchargements des autres utilisateurs, et suivant le caractère plus ou moins saillant de cette information (les titres étaient classés en fonction de leur nombre de téléchargements ou affichés dans un ordre aléatoire). Les auteurs ont ainsi obtenu trois degrés différents d'influence sociale : nulle, faible et forte. Enfin, chaque condition expérimentale a donné lieu à plusieurs expérimentations avec des cohortes d'utilisateurs indépendantes, ce qui a permis de tester la variabilité des évolutions relatives aux différents protocoles.

Bien que les auteurs de cette étude n'aient vraisemblablement pas eu connaissance de la conjecture de von Foerster, leurs conclusions[iii] résonne singulièrement avec celles de son auteur :

*[...] our findings suggest that social influence exerts an important but counterintuitive effect on cultural market formation [...] On the one hand, the more information participants have regarding the decisions of others, the greater agreement they will seem to display regarding their musical preferences; thus the characteristics of success will seem predictable in retrospect. On the other hand, looking across different realizations of the same process, we see that as social influence increases, which particular products turn out to be regarded as good or bad becomes increasingly unpredictable, whether unpredictability is measured directly or in terms of quality.*

L'intuition de von Foerster trouve ici une illustration : le renforcement et l'apparente autonomisation des comportements collectifs avec l'accroissement des influences interpersonnelles ; la transformation de contingences en « destin ».

Mais nous n'avons là qu'une seule face de la médaille, le point de vue intérieur au système. Le tableau a été complété récemment par un article adoptant un point de vue extérieur (Ormerod and Glass 2010). A partir des données produites au cours des expériences de (Salganick et al. 2006) (l'ensemble des traces numériques des participants à ces expériences), les auteurs ont adopté le point de vue extérieur du *data scientist*, qui cherche à prévoir l'évolution du système à partir de l'observation exhaustive de ses états successifs. Le participant et sa vision limitée du système a ainsi laissé place à l'agent omniscient disposant à tout moment l'historique exhaustif des états du système. Encore une fois, les conclusions des auteurs ont une similarité frappante avec les intuitions de von Foerster. Les auteurs ont ainsi montré qu'en situation d'influence sociale forte, une fois qu'une expérience est entamée, il était possible de prédire, au vu d'une petite partie de l'historique des téléchargements, quel allait être son résultat. Alors qu'ils constatent que « social influence decreases the *ex ante* predictability of the ensuing social dynamics » ils montrent que « these same social forces can increase the extent to which the outcome of a social process can be predicted very early in the process », ce qui est exactement la deuxième partie de l'intervention de von Foerster.

Les conclusions de ces deux études empiriques mises en vis-à-vis constituent, plus de trente ans après la formulation de la conjecture, une première prédiction du théorème de von Foerster confirmée expérimentalement et mettant en avant les conséquences de l'accroissement des influences interpersonnelles dans les relations humaines - ou pour reprendre les termes de von Foerster, de la « rigidité ». Du point de vue des acteurs, intérieurs au système, les dynamiques sociales semblent s'autonomiser par rapport aux agents qui les engendrent (émergence de comportements collectifs saillants, imprédictibilité *ex ante* des réalisations) ; alors que d'un point de vue extérieur, les dynamiques sociales apparaissent au contraire comme plus facilement prédictibles.

**L'éclairage de la cybernétique de second ordre**

Il n'est pas anodin que les résultats de Salganick et al. insistent sur le rôle de l'influence sociale dans le processus de formation des préférences, interprétation qui soit dit en passant, va à l'encontre des hypothèses en vigueur dans la plupart des travaux de l'économie néoclassique. La qualité d'un morceau de musique est-elle une propriété du morceau, une propriété du medium qui le diffuse ou une construction sociale ? Bien que leur étude n'ait pas pour but de répondre explicitement à cette question, elle suggère néanmoins que la distribution des préférences exprimées pour tel ou tel titre n'est pas la simple expression d'une valeurs objective. Elle dépend de la configuration des interactions, médiatisées dans cette expérience par une plateforme en ligne.

La cybernétique de second ordre nous permet de penser ce phénomène à travers une conception constructiviste de la cognition. La définition la plus commune aujourd'hui, issue du cognitivisme, envisage la cognition comme une manipulation de représentations portant sur les objets de notre environnement. Pour von Foerster au contraire, il n'y a pas d'environnement objectif en dehors de la cognition. Il définit celle-ci comme l'émergence d'activités neuronales propres à l'observateur, appelés *eigenbehaviours*[iv] (*comportements propres*), résultant de son interaction avec l'environnement (Von Foerster 1981). Ces activités nerveuses sont perçues de manière interne comme comme des pensées et de la volonté, ou de manière externe comme du langage ou du mouvement (Varela 1981). Par conséquent, les objets n'ont pas de propriété objective ; est objet à nos yeux tout ce qui peut résonner avec des comportements propres que *nous* pouvons établir[2] (Clarke et al. 2009).

La conjecture de von Foerster est donc étroitement liée à la question bien plus générale et fondamentale de l'origine des valeurs que nous attribuons aux différents objets et situations que nous rencontrons. Deux conceptions opposées dominent les sciences sociales selon que ces valeurs sont pensées comme des propriétés intrinsèques aux objets ou comme le reflet de normes sociales qui transcendent les individus. Le moteur des dynamiques sociales est alors recherché soit au sein des individus, dans des motivations et préférences qui préexistent à leur participation à la société, soit au contraire au sein d'institutions et de faits sociaux qui contraignent voire déterminent le comportement des individus. C'est l'opposition classique entre individualisme méthodologique, au centre de la plupart des approches en économie, holisme, point de vue largement adopté en sociologie.

Une voie médiane est cependant défendue dans diverses branches des sciences sociales : la sociologie tardienne (Tarde 1893), l'antropologie girardienne (Girard), l'économie des conventions (Orléans 2011) ou l'économie des institutions (Dolfsma 2004) pour n'en mentionner que quelques unes. Toutes ont en commun d'affirmer que les traits des agents *(selon les disciplines : valeurs, motivations, désirs ou préférences, etc.)* sont co-construits au cours de leurs interactions sociales, cette co-construction pouvant être influencée par des prédispositions individuelles ou des structures sociales pré-existantes.

---
2 "present tokens for eigenbehaviors which *we* can establish"

La conception de la cognition prônée par la seconde cybernétique s'inscrit dans cette voie médiane avec ce qui a été appelé *individualisme méthodologique complexe* (Dumouchel and Dupuy 1983) : les sociétés sont des systèmes complexes au sein desquels les configurations et la qualité des interactions révèlent autant qu'elles contraignent les comportements propres (*eigenbehaviors)* des êtres cognitifs qui les composent. Ces *eigenbehaviors* correspondent, en fonction du type de leur manifestation et de leurs échelles de temps caractéristique aux objets étudiés par les sciences sociales : comportements, valeurs, préférences, motivations des agents, etc.

D'où l'importance de la notion de rigidité/trivialité des relations : lorsque la qualité et la quantité des interactions inter-individuelles est modulée, l'essence même des individus qui composent le tissu social est susceptible d'être modifiée. Von Foerster avait d'ailleurs défini le concept de trivialisation quelques années auparavant (von Foerster 1972) comme une intervention sur un système visant à instaurer une relation bijective entre ses entrées (stimulus, causes) et ses sorties (réponses, effets)[v] dans le but de le rendre prédictible. Il était assorti de cette mise en garde qui prend tout son sens ici: « w*hile our pre-occupation with the trivialization of our environment may be in one domain useful and constructive, in another domain it is useless and destructive. Trivialization is a dangerous panacea when man applies it to himself.* »

En affirmant que les comportements (et les préférences) des membres d'une société sont plus ou moins différenciés (distributions concentrées sur quelques valeurs ou étalées) selon que les liens entre individus sont plus ou moins triviaux, la conjecture de von Foerster est avant tout une prédiction portant sur la relation entre les dynamiques des valeurs socio-culturelles et la nature des relations sociales. Dans l'épistémologie de von Foerster, trivialiser les relations entre les hommes revient à trivialiser les êtres eux-mêmes.

L'épistémologie de la seconde cybernétique change donc inévitablement la nature des questions que soulèvent les TIC. Si l'on reprend l'exemple de Salganick et al., la question n'est plus, comme bien souvent dans ce type d'étude, « quelle type d'infrastructure permettra aux utilisateurs de découvrir les meilleurs morceaux étant données leurs préférences ?», mais plutôt « quelles influences ont les différents types d'infrastructures sur les préférences et de quelle manière les *nouvelles* préférences seront satisfaites ?». Il ne s'agit pas là de soutenir une détermination des préférences par les infrastructures techniques, mais de reconnaître d'une part, le caractère changeant des préférences, et d'autre par l'enchevêtrement des volontés individuelles et des configurations d'interactions sociales (médiatisées ou non par la technique) dans les processus de formation des préférences.

Cette dernière question se pose de manière aiguë pour toutes les infrastructures techniques qui régissent nos rapports à nos pairs et à la société. Les mécanismes de recommandation, et en particuliers ceux implémentés par les réseaux sociaux, sont particulièrement concernés du fait qu'ils étendent les possibilités de trivialisation des relations humaines en offrant des jugements et des actions dont la facilité d'intégration ou de mise en œuvre sont proportionnelles à leur standardisation (rating, likes, etc.). Ces effets sont d'autant plus importants qu'ils introduisent de nouvelles rétroactions positives du tout vers les parties.

Inévitablement, par leur omniprésence, les technologies de l'information et de la communication (TIC) induisent des changements sociaux qui touchent à la nature même des dynamiques sociales.

## 5. Vers des sciences sociales prédictives

Des « flash mobs » (mobilisations éclair) aux révolutions arabes de 2011, l'avènement des TIC dans notre quotidien a déjà profondément modifié nos sociétés dans le sens de l'intuition de von Foerster, renforçant l'interdépendance entre les individus bien au-delà des cercles traditionnels, rendant ainsi possible de nouveaux types de comportements collectifs auto-organisés. Mais parallèlement, l'existence d'un point de vue « extérieur » aux interactions humaines, inconcevable à l'ère pré-Internet de von Foerster, ne cesse de se concrétiser à mesure qu'un nombre croissants d'activités humaines laissent des traces numériques (ex. échanges boursiers, e-commerce, publication en ligne, etc.) et que les utilisateurs des nouveaux services apportés par les TIC renoncent à des pans entiers de le vie privée. Que ce soit dans les domaines scientifique, économique, politique, culturel ou industriel, la collecte à grande échelle des traces numériques sociétales est au cœur d'importants enjeux, devenant l'un des secteurs les plus florissants de l'économie numérique.

A l'ère du numérique, peuvent en effet être analysés de manière massive les écrits (ex. mails, blogs, presse, sms, etc.), les données audio (grâce aux progrès de la reconnaissance vocale et la téléphonie via Internet), les centres d'intérêt (ex. analyse des requêtes dans les moteurs de recherche, profils de navigation web, profils d'utilisateurs), les réseaux familiaux, d'amitié, de rencontre et réseaux professionnels (via l'analyse des carnets d'adresse et réseaux sociaux en tout genre), les lieux de vie privilégiés et les déplacements (ex. données GPS ou téléphonie mobile), les images (grâce à la reconnaissance faciale et aux méta-données il est possible de savoir qui est avec qui, quand et où). Cette liste est loin d'être exhaustive et ces données sont de moins en moins considérées comme personnelles par leurs propriétaires. Les droit d'accès aux données privées que requiert une simple smartphone 'app' d'un transporteur public (en l'occurrence RATP Paris version 2014) a de quoi faire réfléchir sur l'érosion de la notion de « vie privée ». Pour bénéficier d'un service aussi simple que la suggestion d'itinéraires de métro, l'utilisateur doit donner accès au statut et à l'identifiant du téléphone, à sa position GPS, à son carnet d'adresse, au stockage de données (lecture et écriture), à tous les réseaux et doit donner les droits de créer ou d'effacer des comptes utilisateurs, de contrôler le vibreur et le système de veille du téléphone, ainsi que les paramètres de synchronisation.

Dans ce contexte, on ne s'étonnera pas que la question de la prédiction de comportements et de la prédiction sociale soit devenue l'un des nouveaux Eldorado de l'économie numérique. Alors qu'IBM vend des systèmes de prédiction des crimes aux polices municipales américaines, Google s'est offert la start-up de prédiction sociale Behavio. Ses services comme *Google Now* se donnent pour objectif de prédire vos comportements pour mieux vous informer. De son côté, Amazon a déposé en 2014 un brevet sur la livraison prédictive (soyez livrés avant même de commander). Tandis que recherche académique ou privée se met à imaginer ce que serait un dispositif de « météo sociale », nous assistons à une effervescence de travaux portant sur la valorisation des traces numériques à des fins prédictives. Les marchés culturels sont bien sûr abordés (Mishne et Glance 2006), la question de la prédiction de blockbusters, dans la vie réelle cette fois, se retrouvant par ailleurs en bonne place comme démonstrateur pour le service de prédiction à la demande de Google. Mais on trouve également la question de la prédiction des cours de la bourse (Bollen et al. 2011), des résultats d'élections (Tumasian et al 2010), des crimes[vi] (Vlahos 2012) ou d'événements tels que les changements de régimes politiques (Leetaru 2011).

Si la quête d'un point de vue « extérieur » à nos sociétés est devenue un enjeu stratégique pour la plupart des grandes entreprises du Web, les récentes révélations de l'ex-agent de la NSA Edward Snowden lui apportent une dimension politique (Greenwald 2014). La nouvelle posture de la NSA « Partner it All, Sniff it All, Know it All, Collect it All, Process it All, Exploit it All » exprimée dans l'un des documents top-secrets qu'il a divulgués laisse peu de doutes sur les ambitions de l'agence. En espionnant les principales autoroutes de l'information à l'échelle planétaire, tout en négociant avec les grandes entreprises des TIC un accès exhaustif aux données de leurs utilisateurs (Microsoft, Google, Apple, Facebook, Yahoo !, Skype, etc.), la NSA s'est constitué une base de données jusqu'alors inimaginable sur l'état des activités humaines, mise à jour quasiment en temps réel. Cette base de données et les méthodes qui seront développées pour l'exploiter constituent peut-être l'horizon ultime de ce que pourrait être un point de vue « extérieur » sur nos sociétés. L'ampleur de l'espionnage de la NSA a choqué par le fait qu'il annihilait toute notion de vie privée autant pour les résidents américains que pour le reste du monde et sans discrimination. Mais le plus troublant est peut-être que les moyens employés semblent disproportionnés, voir inadéquats, au regard de leur justification. La NSA dit nécessiter toutes ces données pour prévenir un second 9/11. Dans le but de trouver une aiguille dans une botte de foin, l'accès à l'ensemble de la botte de foin serait nécessaire. Mais cette affirmation a été sérieusement remise en question, notamment par un rapport de 2013[vii] commandé par la Maison Blanche à un panel d'experts, intitulé « Report of the Review Group on Intelligence and Communications Technologies ». Celui-ci affirme que ce programme de anti-terroriste de la NSA pour la collecte et l'analyse des traces numériques produites à l'échelle mondiale, « was not essential to preventing attacks and that much of the evidence it did turn up "could readily have been obtained in a timely manner using conventional [court] orders." 

Si elles ne le sont pas déjà, la détection de signaux faibles et la question de la prédiction sociale pourrait très bien devenir pour la NSA, par dépit ou par opportunisme, l'une des principales finalités d'une telle entreprise de collecte massive de données. Les enjeux dépassent de loin la lutte contre le terrorisme et sont dans un sens beaucoup plus stratégiques. Quel État ne serait pas intéressé par le fait de prévoir, à l'étranger comme sur son territoire, les prochaines révolutions ou les futurs mouvements de contestation, les prochaines crises économiques ou ce à quoi les électeurs seront le plus réceptifs lors de la prochaine élection ? Si de grandes entreprises du Web, qui ne possèdent pourtant qu'un point de vue limité comparativement à la NSA, voient la prédiction sociale comme un axe stratégique, il serait étonnant qu'il n'en soit pas de même pour une telle institution. La prédiction (en probabilité) des comportements collectifs est par ailleurs, d'un point de vue méthodologique, un usage beaucoup plus adéquat de ces données massives que la détection de potentiels terroristes. Il est en effet probable que taux d'erreur lors de l'identification de potentiels terroristes par ces techniques soit supérieur de quelques ordres de grandeur à proportion de la population ciblée."

Mais dans la mesure où la possibilité de la prédiction sociale dépend de la rigidité des relations humaines, leur domaine d'application se limite aux activités humaines suffisamment trivialisées pour être immunisées contre le changement, comme le soulignait von Foerster il y a plus de quarante ans (Von Foerster 1972):

*In order to protect society from the dangerous consequences of change, not only a whole branch of business has emerged, but also the Government has established several offices that busy themselves in predicting the future by applying the rules of the past. These are the Futurists. Their job is to confuse quality with quantity, and their products are "future scenarios" in which the qualities remain the same, only the quantities change: more cars, wider highways, faster planes, bigger bombs, etc. While these "future scenarios" are meaningless in a changing world, they have become a lucrative business for entrepreneurs who sell them to corporations that profit from designing for obsolescence.*

Par ailleurs, l'élaboration de macroscopes (Rosnay 1975) permettant d'accéder à un point de vue "extérieur" à une société modifie inévitablement, par interférence, la nature du système observé, que ce soit parce que les utilisateurs de ces outils en font partie ou parce que les autres membres de la société se savent observés. Cela soulève des questions éthiques qu'il ne faut pas négliger. Comme en témoigne Edward Snowden (Poiras 2015), les violations répétées de la vie privée des Internautes ont modifiées leur usage d'Internet, leurs comportements en ligne étant bien différents du temps où cet espace était perçu comme une agora mondiale pour l'échange entre les peuples et la libre expression. Dans le domaine des affaires humaines, les efforts visant à atteindre un « point de vue extérieur » modifient la forme de l'objet qu'ils visent. Les révélations Snowden, par leur ampleur, risquent fort d'apporter ce que les experts en prédiction sociale redoutent le plus : le changement.

On peut néanmoins conjecturer que l'introduction de macroscopes, qui ne manquera pas de se démocratiser dans les prochaines années, aura pour conséquences de renforcer la stabilité des systèmes sociaux (attracteur plus stables, effets de polarisation plus marqués) tout en augmentant l'amplitude des phénomènes extrêmes (changement sociaux plus radicaux impliquant une proportion plus importante de l'humanité).

**6. La conjecture de von Foerster en pratique**

La migration de certaines activités sociales vers des supports numériques, par l'ajout de nouvelles dimensions aux liens inter-personnels, a révélé le caractère visionnaire de la conjecture de von Foerster. Elle instaure également de nouveaux rapports entre la collectivité humaine et certains de ses composants.

Alors qu'il y a quelques années le Web pouvait paraître aux yeux de certains comme un univers virtuel, relativement décorrélé du monde réel, le monde numérique fait aujourd'hui partie des territoires que nous arpentons chaque jour. Il s'y déroule une part importante de notre vie sociale, en tant qu'émetteur et récepteur d'information ou membre de réseaux sociaux. Ce nouveau territoire n'en garde pas moins les propriétés de son support, à savoir son artificialité. Dans la dialectique des points de vue intérieur et extérieur, se pose alors la question de la part de réalité exprimée par le monde numérique et l'influence d'artefacts créés dans le but d'influencer le cours de son histoire.

Alors que cela été difficilement imaginable à l'époque von Foerster, les TIC permettent aujourd'hui à une entité quelconque, par une démultiplication artificielle, d'agir sur les dynamiques sociales en profitant de la rigidité des liens inter-personnels. Encore une fois, le secteur du marketing fut pionnier dans cette aventure. La création de faux commentaires par de faux utilisateurs sur tel ou tel produit pour augmenter leurs ventes est désormais pratique courante, bien qu'illégale dans de nombreux pays.

Cette pratique s'est rapidement démocratisée bien au-delà du secteur de la consommation. Des services en ligne proposent aujourd'hui, pour quelques dizaines de dollars, 10.000 'likes' ou followers supplémentaires pour votre profil sur un réseau social lambda. L'industrie des réseaux sociaux commence d'ailleurs à réagir à ce genre de pratiques en « purgeant » de temps à autre de sa base les utilisateurs aux comportements factices. On a pu voir ainsi en 2014 sur Instagram la plupart des utilisateurs perdre une grande partie de leurs abonnés, jusqu'à plusieurs millions pour certaines célébrités, générant une épidémie de crises existentielles bien réelles[viii].

Même les institutions censées réguler ces pratiques y trouvent un intérêt. Alors que la préparation à la cyber-guerre figure à l'agenda de la plupart des grandes puissances mondiales, des services se créent ici ou là pour influencer les opinions sur le Web par la création d'utilisateurs factices - « sock puppets » dans le jargon. Ainsi, le Guardian révélait en mars 2011 (Fielding and Cobain 2011) un appel d'offre de l'armée américaine portant sur des logiciels capables de créer de multiples personas, c'est-à-dire de faux utilisateurs ayant un historique d'activité en ligne (postes de blog, commentaires, tweet, etc.) et une présence dans le cyber-espace qui soit suffisamment cohérents d'un point de vue technique, géographique et culturels pour passer pour des humains. Cette armée de faux utilisateurs, commandée par quelques individus outillés de ces logiciels de gestion de « personas », est destinée à s'insérer dans de multiples réseaux sociaux et plate-formes du web 2.0, et utiliser le moment venu sa capacité d'influence pour diffuser les idées appropriées aux objectifs recherchés (discréditer un individu, faire passer une réforme, soutenir un candidat, etc.).

Officiellement développée pour manipuler l'opinion en pays ennemis, l'armée américaine est soupçonnée d'utiliser cette technologie à l'encontre de ses propres citoyens, comme cela a été le cas dans l'affaire de la rupture des digues provoquée par l'ouragan Katrina, pour discréditer les collectifs qui mettaient en cause la négligence des ingénieurs de l'armée[ix]. D'autres gouvernements utiliseraient également ce type de technologies, comme par exemple le gouvernement iranien, qui se targue d'avoir la seconde cyber-armée au monde, et dont le cyber département Paydari infiltre Facebook pour engager des discussions pro-gouvernementales avec d'autres utilisateurs. Alors que la propagande traditionnelle déployait d'importants efforts pour faire passer des messages via des individus réputés, le cyber-espace ouvre la voie à une stratégie complémentaire. Prenant au mot la vision keynésienne selon laquelle « on ne peut avoir raison contre la foule » (Orléan 1986), il s'agit de donner l'illusion d'une foule parlant d'une même voix pour mieux convaincre.

Cette pratique consistant à créer l'illusion d'un consensus ou du soutien spontané d'une population à une cause a reçu l'appellation anglo-saxonne « d'astroturfing », du nom d'une marque de gazon artificiel. Bien qu'elle existait déjà dans le domaine du marketing, du lobbying industriel ou de la politique avant l'arrivée des réseaux sociaux sur Internet, elle a pris ces dernières années des proportions telles qu'elle a pesé régulièrement sur des décisions politiques importantes ou des résultats électoraux. Par exemple, des chercheurs dans le domaine des systèmes complexes et de la fouille de données ont développé récemment des techniques de reconstruction des flux d'information sur la plate-forme de micro blogging Twitter permettant de détecter des activités d'astroturfing. Ils ont ainsi mis en évidence plusieurs cas de manipulation d'opinion par l'utilisation de faux comptes Twitter dans le cadre d'élections aux États-Unis (Ratkiewicz et al. 2010). Leur plate-forme « Truthy » (http://truthy.indiana.edu), en fournissant à tout internaute une reconstruction des dynamiques informationnelles sur Twitter, permet d'en identifier les usages déviants et contribue à déjouer les pratiques d'astroturfing.

Ce type de macroscope peut contribuer à limiter les effets des pratiques d'astroturfing. Mais dans la mesure où celles-ci gagnent en efficacité à mesure que les liens sociaux se rigidifient, c'est avant tout l'éducation des citoyens qui permettra d'en limiter les effets : apprendre à sélectionner ses sources d'information tout en ayant un point de vue critique sur celles-ci et les recommandations qui en émanent.

**7. Conclusion**

En suivant le fil d'Ariane esquissé par von Foerster il y a plus de 30 ans, nous avons pu entrevoir les modifications profondes qu'a subi notre environnement informationnel ces dernières années. Si les TIC ont pu susciter l'espoir d'un monde commun et pluraliste, le respect des principes démocratiques dans l'espace public numérique est loin d'être acquis et requiert de mieux connaître les lois qui en régissent ses infrastructures.

Tout d'abord en prenant conscience que la manière dont ces espaces sont peuplés d'expression publique est fortement contrainte par leur design et les fonctionnalités qu'ils offrent, tels que les systèmes de recommandation. Ces derniers ont non seulement le potentiel d'influencer les dynamiques socio-économiques à travers les changements des connaissances des agents (problème que Hayek a qualifié de « division of knowledge ») ; mais également d'influencer la formation des préférences au sein d'une population. Ensuite en prenant la mesure du changement d'échelle induit par l'extension de la circulation d'information et l'accroissement de la coordination entre les individus. La société tend à fonctionner de plus en plus comme un système intégré, dont les leviers d'action permettront à qui sait s'en servir d'avoir un impact significatif sur son évolution.

Dans ce contexte, alors que la course à la reconstruction et à la prédiction des dynamiques sociales est ouverte, la possibilité d'une intervention massive de « bas niveau » sur les systèmes sociaux ou au contraire l'accès exclusif à un point de vue « extérieur » sur la société conférera à ceux qui en auront la jouissance un pouvoir de domination contraire aux principes éthiques les plus fondamentaux (voir par exemple Epstein et Robertson 2015 pour une illustration empirique). Le théorème de von Foerster nous donne des pistes pour atténuer ces effets, par exemple en implémentant des systèmes de recommandation qui s'appuient sur des personnes connues dans le monde « réel » et en éduquant les populations à choisir de manière plus réfléchie les degrés de rigidité de leurs relations sociales.

Mais avant tout, nous devons considérer l'éthique qui encadre les recherche de von Foerster. Celle-ci caractérise une société en bonne santé par le fait que chacun de ses membres perçoit l'autre comme un être non trivial (von Foerster 1972). Dans une telle société "l'éducation n'est ni un droit, ni un privilège, mais une nécessité". Chacun mesurera le chemin à parcourir.

Néanmoins, le fait que des organismes tels que la NSA puissent mener conjointement des programmes destinés à constituer un point de vue « extérieur » au sens de von Foerster et des programmes qui, des mots mêmes de la NSA (Greenwald 2014), ont pour objectifs d'utiliser les technologies de du Web pour provoquer quelque chose dans le monde réel ou le cyberespace[3] via « *the 4 D's : Deny/Disrupt/Degrade/Deceive* », doit nous inciter à la plus grande vigilance concernant le développement de leurs capacités à influencer les dynamiques collectives. Ce genre de programmes menacent non seulement les droits fondamentaux, violant sans discrimination la vie privée de la population mondiale, mais également le fonctionnement démocratique de nos sociétés en ouvrant la voie à une *ingénierie sociale* capable d'en détourner les mécanismes. Avec l'avènement du Big Data et le perfectionnement inédit des techniques de collecte et de traitement des données, tout porte à croire qu'en matière de renseignement nous nous approchons du seuil de contre-productivité théorisé par Illich (Illich 1973) : « Lorsqu'une activité outillée dépasse un seuil défini par l'échelle *ad hoc*, elle se retourne d'abord contre sa fin, puis menace de destruction le corps social tout entier. »

**Bibliographie**

---

3 «*make something happen in the real or cyber world* »

[i] Dans le cas d'une société, la *contre-productivité* caractérise un système qui échappe au contrôle des hommes y prenant part et est détruit par les moyens mêmes qui sont censés le servir, tel un organisme frappé d'une maladie auto-immunitaire : « la médecine corrompt la santé, l'école bêtifie, le transport immobilise, les communications rendent sourd et muet, les flux d'information détruisent le sens, [...] l'alimentation industrielle se transforme en poison » [DUP 06].

[ii] Étude de la compagnie Nielsen sur 25.000 internautes, Juillet 2009, http://blog.nielsen.com/nielsenwire/wp-content/uploads/2009/07/pr_global-study_07709.pdf

[iii] On remarquera que ce type d'étude n'a été rendu possible que très récemment grâce aux nouvelles possibilités qu'offrent les plates-formes en ligne, notamment au regard de la taille des cohortes.

[iv] On peut en première approximation comparer les *eigenbehaviours* aux modes propres d'une corde d'instrument de musique. Seuls les sons comprenant des harmoniques de la note fondamentale de la corde la feront résonner. Les eigenbehaviors du cerveaux sont bien plus complexes en ce sens que celui-ci comporte un nombre indéfini de « cordes » inter-corrélées, en résonance avec des multiples dimensions sensorielles, qui de plus se créent et s'ajustent à travers des processus d'apprentissage.

[v] a «one-to-one relationship between its " input " (stimulus, cause) and its " output " (response, effect)»

[vi] On pourra mentionner notamment, dans un registre relativement proche, le développement chez IBM de méthodes d'analyse prédictive à partir de grandes masses de données fournies en temps réel. Celles-ci ont notamment été utilisée, apparemment avec succès, par la police de Memphis pour déployer au mieux leurs effectifs à partir d'une carte de la ville donnant les probabilités d'occurrence de crimes pour les prochaines heures. Ainsi on peut lire dans les supports de communication d'IBM : *« Predictive analytics gives government organizations worldwide a highly-sophisticated and intelligent source to create safer communities by identifying, predicting, responding to and preventing criminal activities. It gives the criminal justice system the ability to draw upon the wealth of data available to detect patterns, make reliable projections and then take the appropriate action in real time to combat crime and protect citizens. »*

[vii] Voir notamment l'article du Washington Post « Officials' defenses of NSA phone program may be unraveling » par By Greg Miller and Ellen Nakashima December 19, 2013, http://www.washingtonpost.com/world/national-security/officials-defenses-of-nsa-phone-program-may-be-unraveling/2013/12/19/6927d8a2-68d3-11e3-ae56-22de072140a2_story.html

[viii] Voir par exemple l'article du *New Yorg Magazine* http://nymag.com/thecut/2014/12/insta-rapture-caused-an-existential-crisis.html

[ix] Voir par exemple l'émission d'investigation de 4WWL broadcastée sur YouTube : http://www.youtube.com/watch?v=12_tsowgA9Q